\documentclass[14pt,a4paper]{article}

\usepackage[english]{babel}
\usepackage{amsmath,amssymb,amsfonts,amsthm}
\usepackage[mathscr]{eucal}
\usepackage[all]{xy}
\usepackage{hyperref}
\usepackage{setspace}
\usepackage{upgreek}
\usepackage{hyperref}

\usepackage{times}
\usepackage{color}

\usepackage{indentfirst}

\begin{document}

\title{Stable Interactions between Higher Derivative Extended Chern-Simons and Charged Scalar Field}
\author {V.~A.~Abakumova\footnote{abakumova@phys.tsu.ru}, \ D.~S.~Kaparulin\footnote{dsc@phys.tsu.ru}, \ and S.~L.~Lyakhovich\footnote{sll@phys.tsu.ru}}
\date{\footnotesize\textit{Physics Faculty, Tomsk State University, Tomsk 634050, Russia }}

\maketitle

\begin{abstract}
\noindent We consider inclusion of interactions between the higher derivative extended Chern-Simons and charged scalar field. We demonstrate that the order $N$ extended Chern-Simons and order $2n$ charged scalar admit the $(N+n)$-parameter series of interaction vertices. The interactions are in general non-Lagrangian, but they preserve a certain conserved second-rank tensor, whose parameters are determined by the coupling. The $00$-component of this tensor can be bounded even if the canonical energy of the model is unbounded before the inclusion of interaction. If the $00$-component of conserved tensor is bounded, the theory is stable.
\end{abstract}

\section*{Introduction}

Higher derivative theories are important in modern theoretical physics. The well-known examples include the Pais--Uhlenbeck oscillator \cite{PU50PR}, Podolsky electrodynamics \cite{P42PR}, conformal field theories \cite{FT85PR}, modified theories of gravity \cite{BL87CQG, SF10RMP, FT10LRR, LP11PRL}.  In many cases, the higher derivative models have remarkable properties. They often admit a wider symmetry compared to their counterparts without higher derivatives and have better convergency properties both at the classical and quantum level.

A notorious difficulty of higher derivative models concerns instability of their dynamics. The classical instability is connected with the fact that the canonical energy is typically unbounded for the higher derivative Lagrangians. The quantum instability reveals itself by ghost poles in the propagator and leads to the unbounded energy spectrum. The stability of higher derivative dynamics has been studied for decades \cite{W15S, P16IJGMP, S17IJMPA}. 

The recent observations have shown that higher derivative theory can be stable, even if the energy is unbounded \cite{T15IJMPA, S09SIGMA, PP13LA}. Also for some exceptional class of higher derivative theories the stability is related to the fact that the canonical energy is bounded because of strong second-class constraints \cite{S05NPB, BKRTY11CQG}. Another idea was proposed in \cite{KLS14EPJC}. It has been demonstrated that the wide class of higher derivative theories admits another bounded conserved quantities that stabilize the dynamics.

The stability problem is most relevant at the interacting level. In particular, in nonlinear theory the collapsing trajectories can appear even if the free model is stable \cite{P16IJGMP, S17IJMPA}. This happens because the bounded conserved quantity does not exist at the interacting level. The methods of inclusion of stable interactions to the theory that preserve some special conservation law have been developed in \cite{KLS14EPJC, KLS16JPA}. For the use of adiabatic invariants and discussion of other ideas see \cite{BBDW19EPJC} and references therein.

In higher derivative gauge theories the admissible interactions are restricted by both the stability condition and gauge invariance. The general problem of inclusion of interactions between two higher derivative theories, with one of them being gauge invariant, has been considered in \cite{AKL19PRD}. In the present paper we demonstrate how to construct consistent and stable interactions between the higher derivative gauge invariant extended Chern-Simons and charged scalar field.

\section{Setting}

We consider a theory of charged scalar $\phi=\text{Re}\,\phi(x)+\text{Im}\,\phi(x)$ and gauge vector field $\Phi=\Phi_\mu(x)dx^\mu$ on $3d$ Minkowski space with local coordinates $x^\mu,\,\mu=0,1,2,$ and metric $\eta_{\mu\nu}=\text{diag}\,(+1,-1,-1)$, described by the action functional
\begin{equation}\label{SCSSF}
\phantom{\frac12}\displaystyle S[\phi(x), \Phi(x)]=S_{SF}[\phi(x)]+S_{CS}[\Phi(x)]\,,\phantom{\frac12}
\end{equation}
where 
\begin{eqnarray}\label{SF}
\displaystyle S_{SF}[\phi(x)]&=&\frac{1}{2}\int d^3x \prod\limits_{p=0}^{n-1}\phi^\ast
\big(\square+m^2\alpha_p^2\big)\phi\,, \qquad \square=\partial_\mu\partial^\mu\,,\\[3mm]\label{CS}
\displaystyle S_{CS}[\Phi(x)]&=&\frac{m^2}{2}\int d^3x \sum\limits_{q=1}^NA_q\Phi_\mu(W^q\Phi)^\mu\,, \qquad W\equiv m^{-1}\ast d\,.
\end{eqnarray}
Here, $\ast$ and $d$ denote the Hodge star operator and de Rahm differential, respectively; $m$ is a constant with the dimension of mass. The real numbers $A_q,\,q=1,\ldots,N;\,\alpha_p,\,p=0,\ldots,n-1,$ are parameters of the model. The integers $n, N$ determine the highest order of derivatives of the fields $\phi, \Phi$ in the action (\ref{SCSSF}). Without loss of generality, we assume $A_N\neq0$.

The theories (\ref{SF}), (\ref{CS}) are well-known. The theory (\ref{SF}) describes the multiplet of charge scalar fields with different masses. The extended Chern-Simons theory (\ref{CS}) has been proposed in \cite{DJ99PLB}. It corresponds to a reducible representation of the Poincar\'e group. The specifics of the representation is defined by the structure of roots of the polynomial
\begin{equation}\label{Mchar}
\displaystyle M(z)=\sum\limits_{r=0}^{N-1}A_{r+1}z^r,
\end{equation}
where $z$ is a formal variable. If all the roots of (\ref{Mchar}) are different, the mass spectrum of (\ref{CS}) is non-degenerate, and representation is unitary. The irreducible components are massive and massless spin 1 fields. The action of the Chern-Simons operator $W$  on the vector field is determined by the relation
\begin{equation}\label{}
\displaystyle (W\Phi)_\mu dx^\mu=\varepsilon_{\mu\nu\rho}\partial^\nu\Phi^\rho dx^\mu\,,
\end{equation}
where $\varepsilon_{\mu\nu\rho}$ denotes the $3d$ Levi-Civita symbol with $\varepsilon_{012}=1$. 

The free equations of motion for the action functional (\ref{SCSSF}) read
\begin{equation}\label{EqCSSF}
\displaystyle \frac{\delta S}{\delta \phi}\equiv\sum\limits_{p=0}^{n-1}\big(\square+m^2\alpha_p^2\big)\phi=0\,, \qquad \frac{\delta S}{\delta \Phi^\mu}\equiv\frac{m^2}{2}\sum\limits_{q=1}^NA_q(W^q\Phi)_\mu\,.
\displaystyle 
\end{equation}
These equations are invariant under the standard gradient gauge transformation for the vector field $\Phi$,
\begin{equation}\label{GSf}
\displaystyle \delta_\varepsilon\phi=0\,, \qquad \delta_\varepsilon \Phi=\partial_\mu\varepsilon\,,\end{equation}
where $\varepsilon=\varepsilon(x)$ is a gauge parameter, being the arbitrary function of space-time coordinates. The corresponding gauge identity reads
\begin{equation}\label{GIf}
\displaystyle  \partial^\mu\frac{\delta S}{\delta \Phi^\mu}=\partial^\mu\Big(\frac{m^2}{2}\sum\limits_{q=1}^NA_q(W^q\Phi)_\mu\Big)\equiv0\,.
\end{equation}

We are looking for consistent interactions between theories (\ref{SF}), (\ref{CS}), that are not necessarily Lagrangian. By (non)-Lagrangian interaction \cite{KLS13JHEP} we mean a formal deformation of free equations (\ref{EqCSSF}) of the form
\begin{equation}\label{EE}
\displaystyle E\equiv\frac{\delta S}{\delta \phi}+gE^{(1)}(\phi, \Phi)+g^2E^{(2)}(\phi, \Phi)+\ldots=0\,, \qquad  E_\mu\equiv\frac{\delta S}{\delta \Phi^\mu}+gE^{(1)}{}_\mu(\phi, \Phi)+g^2E^{(2)}{}_\mu(\phi, \Phi)+\ldots=0\,,
\end{equation}
where $g$ denotes a coupling constant, and $E^{(1)}, E^{(2)},\ldots$ are bilinear, trilinear, etc. in the fields and their derivatives. The interaction is consistent if all the gauge symmetries and gauge identities are preserved by coupling. This means that the equations (\ref{EECSSF}) should be gauge invariant,
\begin{equation}\label{GS}
\displaystyle \delta_\varepsilon\phi=R\varepsilon\,, \qquad \delta_\varepsilon\Phi_\mu=R_\mu\varepsilon\,,
\end{equation}
and should admit the Noether identity
\begin{equation}\label{GI}
\displaystyle ZE+Z^\mu E_\mu\equiv0\,.
\end{equation}
The gauge symmetry and gauge identity generators $R, R_\mu, Z, Z^\mu$ (\ref{GS}), (\ref{GI}) are obtained by deformation of corresponding free quantities (\ref{GSf}), (\ref{GIf}):
\begin{eqnarray}\label{}
\displaystyle R&=&gR^{(1)}+g^2R^{(2)}+\ldots\,, \qquad R_\mu=\partial_\mu+gR^{(1)}{}_\mu+g^2R^{(2)}{}_\mu+\ldots\,, \qquad \\[3mm]\label{}
\displaystyle Z&=&gZ^{(1)}+g^2Z^{(2)}+\ldots\,, \qquad Z^\mu=\partial^\mu+gZ^{(1)}{}^\mu+g^2Z^{(2)}{}^\mu+\ldots\,.
\end{eqnarray}
It is known that in the non-Lagrangian theories the gauge symmetry and gauge identity generators are not related to each other. So, we allow $Z\neq R,\,Z^\mu\neq R^\mu$ for $g\neq 0$.

As far as stability is in question, we require that some second-rank tensor should be conserved at the interacting level. This is an additional selection rule for admissible interactions. If the $00$-component of such conserved tensor is bounded, the nonlinear theory (\ref{EE}) is stable.

\section{Main result}

In this section, we explicitly construct consistent interactions between higher derivative extended Chern-Simons and charged scalar field.

Let us introduce the extended covariant derivative
\begin{equation}\label{Dmu}
\displaystyle D_\mu=\partial_\mu-i\beta\Phi_\mu-i\sum\limits_{q=1}^{N-1}B_q\Phi^{(q)}{}_\mu\,,
\end{equation}
where the notation is used:
\begin{equation}\label{}
\displaystyle \Phi^{(q)}=(m^{-1}\ast d)^q\Phi\,, \qquad q=1\,,\ldots\,,\, N-1\,,
\end{equation}
and $\beta,\,B_q,\,q=1,\ldots,N-1,$ are some real constants. The action of (\ref{Dmu}) on the scalar field $\phi$ and its complex conjugate $\phi^\ast$ is determined by the rule
\begin{equation}\label{}
\displaystyle D_\mu\phi=\big(\partial_\mu-i\beta\Phi_\mu-i\sum\limits_{q=1}^{N-1}B_q\Phi^{(q)}{}_\mu\big)\phi\,, \qquad D_\mu\phi^\ast=\big(\partial_\mu+i\beta\Phi_\mu+i\sum\limits_{q=1}^{N-1}B_q\Phi^{(q)}{}_\mu\big)\phi^\ast\,.
\end{equation}

The scalar field admits the $n$-parameter series of charge currents,
\begin{equation}\label{jjp}
\displaystyle j_\mu(\beta,B)=\sum\limits_{p=0}^{n-1}\beta_pj^{(p)}{}_\mu\,, \qquad j^{(p)}{}_\mu=i\big(\phi^{\ast(p)}D_\mu\phi^{(p)}-\phi^{(p)}D_\mu\phi^{\ast(p)}\big)\,, \qquad \phi^{(p)}=\prod_{\substack{p'=0\\ p'\neq p}}^{n-1}\frac{m^{-2}D_\mu D^\mu+\alpha_{p'}^2}{\alpha_p^2-\alpha_{p'}^2}\phi\,,
\end{equation}
where $\beta_p,\,p=0,\ldots,n-1,$ are some real constants. The vectors $j^{(p)}{}_\mu,\, p=0,\ldots,n-1$, are the charge currents of the irreducible components $\phi^{(p)}$ of a reducible representation of the Poincar\'e group described by the theory (\ref{SF}). In $j_\mu$, the coefficients $\beta_p$ can be interpreted as the charges of components. The divergence of $j^{(p)}{}_\mu$ (\ref{jjp}) reads
\begin{equation}\label{partialmujmu}
\displaystyle \partial^\mu j^{(p)}{}_\mu=i\prod_{\substack{p'=0\\ p'\neq p}}^{n-1}\frac{1}{\alpha_p^2-\alpha_{p'}^2}\big(\phi^{\ast(p)}\prod\limits_{q=0}^{n-1}\big(D_\mu D^\mu+m^2\alpha_q^2\big)\phi-\phi^{(p)}\prod\limits_{q=0}^{n-1}\big(D_\mu D^\mu+m^2\alpha_q^2\big)\phi^\ast\big)\,. 
\end{equation}
As we see, $j^{(p)}{}_\mu$ is a conserved current if scalar field $\phi$ meet the covariant generalization of free equations (\ref{EqCSSF}).

We introduce the interacting equations between higher derivative extended Chern-Simons and charged scalar field in the following form:
\begin{equation}\label{EECSSF}
\displaystyle E\equiv \prod\limits_{p=0}^{n-1}\big(D_\mu D^\mu+m^2\alpha_p^2\big)\phi=0\,, \qquad E_\mu\equiv\frac{m^2}{2}\sum\limits_{q=1}^NA_q\Phi^{(q)}{}_\mu+j_\mu(\beta,B)=0\,,\end{equation}
where $j_\mu(\beta,B)$ is defined by (\ref{jjp}), and $D_\mu$ is the extended covariant derivative (\ref{Dmu}). The parameters $\beta, \beta_p, B_q,\,p=0,\ldots,n-1,\,q=1,\ldots,N-1,$ are considered as the coupling constants. There are $n+N$ independent coupling parameters in equations (\ref{EECSSF}). The interactions (\ref{EECSSF}) are non-Lagrangian in general. The Lagrangian interaction corresponds to the following values of the parameters:
\begin{equation}\label{betaL}
\displaystyle \beta_0=\beta\,, \quad B_1=\ldots=B_{N-1}=\beta_1=\ldots=\beta_{n-1}=0\,.
\end{equation}

Let us discuss the consistency of interactions. The equations (\ref{EECSSF}) are obviously gauge invariant with respect to the standard $U(1)$-transformation for scalar field and gradient transformation for vector field,
\begin{equation}\label{}
\displaystyle \delta_\varepsilon\phi=i\beta\phi^\ast\varepsilon\,, \qquad \delta_\varepsilon\Phi_\mu=\partial_\mu\varepsilon\,.
\end{equation}
The Noether identity between equations (\ref{EECSSF}) follows from (\ref{GIf}), (\ref{partialmujmu}),
\begin{equation}\label{GIEE}
\displaystyle \partial^\mu E_\mu-i\sum_{p=0}^{n-1}\prod_{\substack{p'=0\\ p'\neq p}}^{n-1}\frac{\beta_p}{\alpha_p^2-\alpha_{p'}^2}\big(\phi^{\ast(p)}E-\phi^{(p)}E^\ast\big)=0\,. 
\end{equation}
The conserved tensor of the interacting theory (\ref{EECSSF}) reads
\begin{equation}\label{ThetaCSSF}
\displaystyle \phantom{\frac12}\Theta_{\mu\nu}(\phi, \Phi)=\Theta_{\mu\nu}^{SF}(\phi,\Phi)+\Theta_{\mu\nu}^{CS}(\Phi)\,.\phantom{\frac12}
\end{equation}
The quantities $\Theta_{\mu\nu}^{SF}(\phi,\Phi)$ and $\Theta_{\mu\nu}^{CS}(\Phi)$ are defined as follows:
\begin{equation}\label{}
\displaystyle \Theta_{\mu\nu}^{SF}(\phi,\Phi)=\sum_{p=0}^{n-1}\beta_p\big[D_{\mu}\phi^{\ast(p)}D_{\nu}\phi^{(p)}+D_{\nu}\phi^{\ast(p)}D_{\mu}\phi^{(p)}-\eta_{\mu\nu}\big(D_\lambda\phi^{\ast(p)}D^\lambda\phi^{(p)}+m^2\alpha_p^2\phi^{\ast(p)}\phi^{(p)}\big)\big]\,;
\end{equation}
\vspace{-3mm}
\begin{equation}\label{}
\displaystyle  \Theta_{\mu\nu}^{CS}(\Phi)=\frac{m^2}{2}\sum\limits_{r,s=1}^{n-1}C_{r,s}(A,B)\big[\Phi^{(r)}{}_\mu \Phi^{(s)}{}_\nu+\Phi^{(r)}{}_\nu \Phi^{(s)}{}_\mu-\eta_{\mu\nu}\Phi^{(r)}{}_\lambda\Phi^{(s)}{}^\lambda\big]\,,
\end{equation}
where $C_{r,s}(A,B)$ is the Bezout matrix of two polynomials, defined by the generating relation
\begin{equation}\label{Crs}
\displaystyle C_{r,s}(A,B)=\frac{\partial^{r+s}}{\partial^rz\,\partial^sz'}\Big(\frac{zM(z)N(z')-z'M(z')N(z)}{z-z'}\Big)\Big|_{z=z'=0}\,, \qquad N(z)=\beta z+\sum\limits_{q=1}^{N-1}B_qz^{q+1}\,,
\end{equation}
where  $z$ and $z'$ are two independent variables, and $M(z)$ is given in (\ref{Mchar}). So, we demonstrated that the interactions (\ref{EECSSF}) satisfy all the consistency conditions.

\section{Stability condition}

In this section, we elaborate on stability of interactions (\ref{EECSSF}). 
The $00$-component of conserved tensor (\ref{ThetaCSSF}) reads
\begin{equation}\label{}
\displaystyle \phantom{\frac12}\Theta_{00}(\phi, \Phi)=\Theta_{00}^{SF}(\phi,\Phi)+\Theta_{00}^{CS}(\Phi)\,,\phantom{\frac12}
\end{equation}
where 
\begin{equation}\label{}
\displaystyle \Theta_{00}^{SF}(\phi,\Phi)=\sum\limits_{\mu=0}^2\sum\limits_{p=0}^{n-1}\beta_p\big(D_\mu\phi^{\ast(p)}D_\mu\phi^{(p)}+m^2\alpha_p^2\phi^{\ast(p)}\phi^{(p)}\big)\,, \qquad  \Theta_{00}^{CS}(\Phi)=\frac{m^2}{2}\sum\limits_{\mu=0}^2\sum\limits_{r,s=1}^{N-1}C_{r,s}(A,B)\Phi^{(r)}{}_\mu\Phi^{(s)}{}_\mu\,.
\end{equation}
The quantity $\Theta_{00}^{SF}$ is a linear combination of bounded conservation laws for interacting components (\ref{jjp}) of scalar field. It can be bounded if $\beta_p>0,\, p=0,\ldots,n-1$\,. The quantity $\Theta_{00}^{CS}$ is a quadratic form in the vector field $\Phi$\,. It is bounded if $C_{r,s}(A,B)$ (\ref{Crs}) is a positive definite matrix. In \cite{KKL15EPJC}, it has been demonstrated that the non-degenerate bounded conserved quantities are admissible by the extended Chern-Simons theory if and only if the polynomial (\ref{Mchar}) has simple and real roots. The corresponding representation of the Poincar\'e group, which is described by the model, includes the multiplet of free self-dual spin 1 fields with different masses, zero mass is admissible. Stable theories transform under unitary representations of the Poincar\'e group. The detailed analysis of stability in the example of third-order extended Chern-Simons theory can be found in \cite{AKL18EPJC}. The stability condition for the interactions (\ref{EECSSF}) reads
\begin{equation}\label{stcond}
\displaystyle C_{r,s}(A,B)\quad \text{is a positive definite matrix}\,,\quad \beta_p>0\,,\, p=0\,,\,\ldots\,,\,n-1\,.
\end{equation}
The Lagrangian interaction vertex (\ref{betaL}) is always unstable. The non-Lagrangian interactions can meet the stability condition for the appropriate choice of coupling constants.

\section*{Conclusion}

We considered the inclusion of consistent interactions between the higher derivative gauge extended Chern-Simons and charged scalar field. The selection rule for the interactions is as follows: the number of gauge symmetries and gauge identities is preserved by the coupling, and at least one conserved second-rank tensor survives at the nonlinear level. It has been demonstrated that the extended Chern-Simons of order $N$ and charged scalar of order $2n$ admit the $(N+n)$-parameter series of interactions (\ref{EECSSF}) with such properties. The conserved tensor is given by equation (\ref{ThetaCSSF}). Its $00$-component can be bounded depending on the values of coupling constants. The conserved tensor with bounded $00$-component stabilizes dynamics of the theory (\ref{EECSSF}). The Lagrangian interaction is unstable, while the non-Lagrangian interactions can be consistent with the stability condition (\ref{stcond}).

\vspace{0.2cm} \noindent
{\bf Acknowledgments.} This research was funded by the state task of Ministry of Science and Higher Education of Russian Federation, grant number 3.9594.2017/8.9. We would like to express our gratitude to the organisers of the AYSS--2019 conference for their hospitality and support.

\end{document}